# Spin dynamics of a solid-state qubit in proximity to a superconductor

Richard Monge[1,2], Tom Delord[1], Nicholas Proscia[1], Zav Shotan[1], Harishankar Jayakumar[1], Jacob Henshaw[1], Pablo R. Zangara[1], Artur Lozovoi[1], Daniela Pagliero[1], Pablo D. Esquinazi[3], Toshu An[4], Inti Sodemann[5,6], Vinod M. Menon[1,2], Carlos A. Meriles[1,2,*]

**A broad effort is underway to understand and harness the interaction between superconductors and spin-active color centers with an eye on the realization of hybrid quantum devices and novel imaging modalities of superconducting materials. Most work, however, overlooks the complex interplay between either system and the environment created by the color center host. Here we use an all-diamond scanning probe to investigate the spin dynamics of a single nitrogen-vacancy (NV) center proximal to a high-critical-temperature superconducting film in the presence of a weak magnetic field. We find that the presence of the superconductor increases the NV spin coherence lifetime, a phenomenon we tentatively rationalize as a change in the electric noise due to a superconductor-induced redistribution of charge carriers near the NV site. We build on these findings to demonstrate transverse-relaxation-time-weighted imaging of the superconductor film. These results shed light on the complex surface dynamics governing the spin coherence of shallow NVs while simultaneously paving the route to new forms of noise spectroscopy and imaging of superconductors.**

As quantum technologies gain momentum, it has become abundantly clear that no physical platform is ideally suited to all practical tasks, a notion presently driving the study of hybrid devices integrating quantum systems with complementary functionalities[1,2]. Among them, much attention is being devoted to physical realizations where an individually-addressable qubit in the form of a trapped ion[3], a cold atom[4], or a molecule[5] interacts with a superconducting qubit or a superconductor electrode in an atom chip[6-12]. Of special interest are all-solid architectures integrating superconductors and adjacent spin-active color centers[13,14] because they can be exploited not only as long-lived quantum memories but also as interfaces, e.g., to convert microwave into optical photons. A closely related research front — sharing the same geometry and ultimately governed by the same physical principles — exploits color centers as local probes[15,16]. Scanning magnetometry of superconductors via nitrogen-vacancy (NV) centers in diamond is a paradigm example that has led to quantitative imaging of individual vortices with exquisite resolution[17-19].

Although color centers in these applications are often seen as isolated systems, the host crystal contains co-existing spins[20-24] as well as fluctuating charges[25-27] whose dynamics are not immune to the superconductor vicinity. Diamond surfaces in particular feature various non-fluorescent paramagnetic centers — e.g., in the form of dangling bonds[28] — as well as shallow carriers, whose type and concentration largely depend on surface termination[29]. Here we use an all-diamond scanning probe to control the distance between an individual tip-hosted NV center and a high-critical-temperature superconductor film. Upon application of pulsed control protocols, we find that close proximity to the superconductor extends the NV spin coherence lifetime. We develop a theoretical formalism that allows us to compare the impact of alternative noise sources near the diamond surface, and propose a superconductor-induced charge rearrangement process as the one responsible for our observations. We then build on the interplay between the superconductor and the NV environment to reconstruct a one-dimensional transverse relaxometry image across the superconductor boundary.

## Results

**Multi-mode scanning imaging of the superconductor film.**

Diamond-based magnetometry builds on the singular properties of the NV center, a spin-1 system amenable to optical initialization and readout via a spin-selective excitation cycle[30]. For sufficiently low magnetic field amplitudes $B_M$, the energies corresponding to the $|m_S = \pm 1\rangle$ eigenstates change linearly with the field projection $B_{NV}$ along the NV symmetry axis. Correspondingly, optically-detected magnetic resonance (ODMR) between one of these two levels and $|m_S = 0\rangle$ — insensitive to $B_{NV}$ — allows one to measure the field projection amplitude. Full vector magnetometry is made possible by comparing the responses from NVs oriented along different crystalline axes[31].

Figure 1a lays out our experimental setup: The system uses confocal microscopy to monitor the fluorescence from an individual NV in an all-diamond probe, in turn connected to the tuning fork of an atomic force microscope (see Supplementary Information (SI), Section I and Figs. S1 and S2). The sample we study is a 500-nm-thick film of $Tl_2Ba_2CaCu_2O_8$ (here referred to as TBCCO), a superconductor material with critical temperature $T_c \approx 105$ K[32]. The film was patterned into square patches of variable size, allowing us to probe superconductivity-induced changes in the NV response as we approach the sample edges (SI, Section II and Fig. S3).

---

[1]Department. of Physics, CUNY-City College of New York, New York, NY 10031, USA. [2]CUNY-Graduate Center, New York, NY 10016, USA. [3]Division of Superconductivity and Magnetism, Felix-Bloch-Institute for Solid State Physics, University of Leipzig, D-04103 Leipzig, Germany. [4]Japan Advanced Institute of Science and Technology, Nomi City, Ishikawa 923-1292, Japan. [5]Institut for Theoretical Physics, University of Leipzig, D-04103 Leipzig, Germany. [6]Max-Planck Institute for the Physics of Complex Systems, D-01187 Dresden, Germany. *E-mail: cmeriles@ccny.cuny.edu

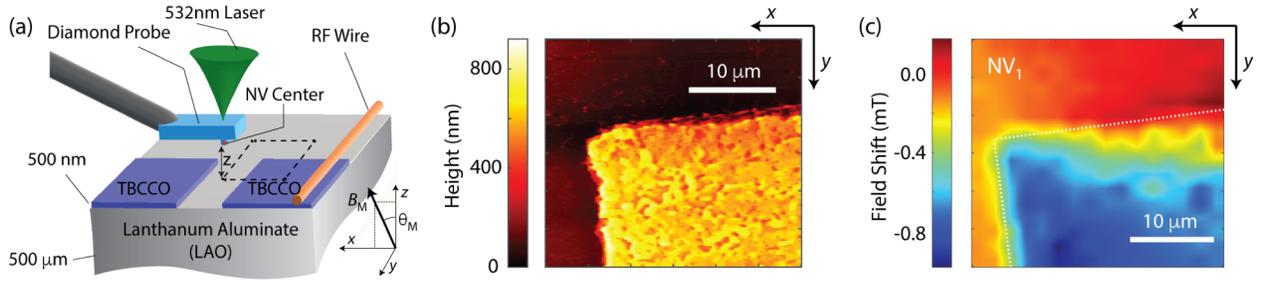

**Fig. 1 | NV magnetometry of a high-critical-temperature superconductor.** (a) We use an NV-hosting scanning probe to monitor a micro-structured TBCCO film patterned into squares of variable size. (b) Topographic image at the corner of a TBCCO square, the region outlined by dashed lines in (a). (c) NV ODMR magnetometry of the same sample section at 69 K in the presence of a uniform magnetic field $B_M \approx 6$ mT at an angle $\theta_M \approx 44$ deg. with the surface normal; the white dashed line indicates the edge of the TBCCO patch. The superconductor shifts the resonance frequency of the probe spin — here dubbed $NV_1$ — as it shields it from the magnetic field. Reduced shielding in the upper right section stems from field realignment near the edge. Here — and everywhere else unless noted — we monitor the $|m_S = 0\rangle \leftrightarrow |m_S = -1\rangle$ spin transition of the NV ground triplet; the tip-TBCCO distance during the magnetometry scan is 300 nm.

We operate our system as an atomic-force or confocal fluorescence microscope, and isolate for inspection a corner in a representative TBCCO patch. Topographic imaging reveals a rugged surface (Fig. 1b), common for this type of sample[33]. On the other hand, Fig. 1c reproduces the NV magnetometry image of the same corner as derived from the frequency shift in the NV fluorescence dip under continuous microwave (mw) excitation (here referred to as "ODMR signal", see SI, Fig. S2). Operating at $T = 69$ K — well below the critical temperature — we observe a reduction of the applied magnetic field on the superconductor film, an indication of (partial) magnetic shielding[15]. Note that the measured profile does not completely match the geometry of the film, a consequence of the varying magnetic field projection on the NV axis. Good agreement, however, can be attained with the help of a second, differently-oriented NV, which also allows us to recover a map of the field orientation (SI, Section III, and Fig. S4).

**NV coherence in proximity to a superconductor**

One of the advantages of NV sensing is the ability to implement time-resolved magnetometry protocols, of interest given the complementary information they convey. Figures 2a and 2b present an example in the form of a Hahn-echo (HE) sequence on the NV spin probe. We observe "revivals" of the echo amplitude stemming from bulk $^{13}$C spin precession at the applied magnetic field[34], whose local value can be extracted from the time interval between them and the known $^{13}$C gyromagnetic ratio. Comparing the NV response above the superconductor or the lanthanum aluminate (LAO) substrate, we observe not only a change in the revival period but also in their relative amplitudes, indicative of a varying spin coherence time. As shown in Figs. 2c and 2d, this change stems from proximity to the superconductor, vanishing as we retract the tip by a few microns. Note that unlike the case in

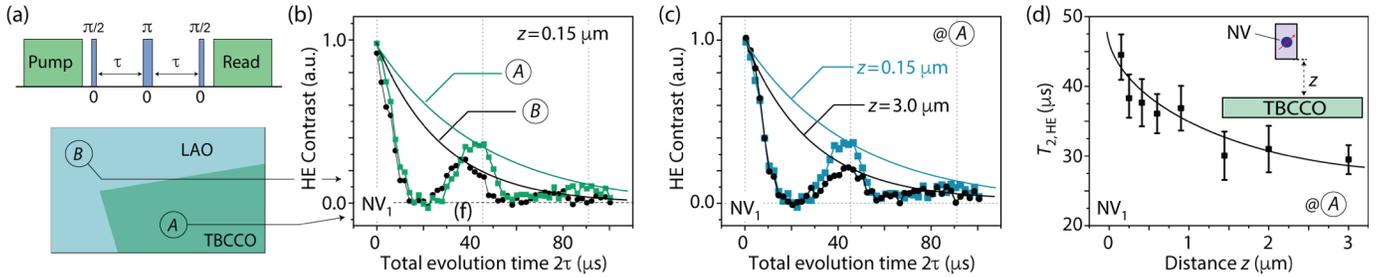

**Fig. 2 | Impact of the superconductor on NV spin coherence.** (a) Pulse protocol; green (blue) blocks indicate laser (mw) pulses on the NV; the upper (lower) labels on the blue blocks indicate the pulse duration (phase). (b) Optically-detected Hahn-echo traces of spin probe $NV_1$ at two different locations, on the superconducting film and away (respectively, points $A$ and $B$ in the left schematic).; squares indicate data points, while solid lines are exponential fits of the envelope. The Hahn-echo times $T_{2,HE}$ are $42.8 \pm 2.9$ μs and $28.3 \pm 3.0$ μs, respectively. (c) Same as before but either close to ($z \approx 0.15$ μm, data in blue) or far from ($z \approx 3.0$ μm, data in black) the surface at point $A$ on the TBCCO film. In (b) and (c), the vertical dashed lines indicate the positions of the echo revivals at point $A$. (d) Hahn-echo coherence lifetime as a function of the distance $z$ between $NV_1$ and the TBCCO surface at point $A$; error bars denote standard deviation and the solid line is a guide to the eye.

Fig. 2b, the revival peak times in Fig. 2c remain unchanged with distance, implying that the magnetic field amplitude is constant within the probed range. Further, we observe comparable fractional $T_2$ changes even in the absence of refocusing pulses — the so-called Ramsey protocol, see SI, Section IV and Fig. S5 — indicating that the TBCCO-induced noise reduction extends to near-zero frequencies.

**Modeling electromagnetic noise near the diamond surface**

Interpreting the above findings is involved since shallow carriers and surface paramagnetic centers create a complex environment featuring alternative electric and magnetic noise mechanisms[21-23,35]. Under the assumption of Gaussian, stationary processes leading to pure dephasing, the noise spectral density $S(\omega)$ relates to the "coherence functional" $\chi(t)$ via the integral equation[36]

$$\chi(t) = -\ln C(t) = \frac{1}{\pi} \int_0^\infty d\omega\, S(\omega)\, \frac{F_n(\omega t)}{\omega^2}. \quad (1)$$

In the above expression, $C(t)$ denotes the in-phase probe spin coherence amplitude after a total time $t = n\tau$, and $F_n(\omega t)$ is a filter function intrinsic to the $n$-pulse protocol[36]. It follows from Eq. (1) that different noise sources contribute to the spin coherence decay with characteristic transverse relaxation times that depend both on the mechanism at play and the control protocol in use.

In order to unravel the system dynamics, we start by considering magnetic interactions of the NV with its environment. Meissner shielding from magnetic fluctuations — arising, e.g., from reconfigurations of paramagnetic centers or transient currents on the diamond tip surface — could hypothetically lead to longer coherences lifetimes, but numerical modeling indicates the effect is marginal (see SI, Section V and Figs. S6 and S7).

The NV center, however, is also sensitive to changes in the distribution of local charges due to the impact of electric fields on the electronic orbitals defining the ground spin triplet[37,38]. In particular, prior work with shallow NVs in bulk crystals has reported longer coherence lifetimes in the presence of liquids with high dielectric constants, suggesting that electric noise from reconfiguring surface charges can play a major role[25,39]. In the present case, the superconductor response can be modeled through a collection of "mirror charges" cancelling the transverse electric field on the TBCCO surface (Fig. 3a). Using a model of Ohmic conduction that accounts for the electrostatic energy cost of charge fluctuations (SI, Section VII), we write the electric noise density from surface carriers at the NV site as

$$S_E(\omega) \sim \left(\frac{\delta_\parallel}{\hbar}\right)^2 \frac{k_B \mathcal{T}}{8\pi} \frac{1}{d^2} \frac{\rho(z)}{\left(1 + \left(\sqrt{8/3}\,\epsilon\omega\rho d\right)^2\right)}, \quad (2)$$

where $\rho$ is the surface resistivity of the diamond tip, $\delta_\parallel$ is the NV–electric field coupling parameter, $\varepsilon$ is the dielectric constant of diamond, and $\hbar$ is the reduced Planck constant.

Eq. (2) shows that $S_E(\omega)$ increases as the system becomes less conducting, precisely the case for oxygen-terminated diamond (where the surface resistivity falls in the range $10^{10} - 10^{17}$ Ohm[40-42]). As shown in Fig. 3b, it also captures the $\omega^{-2}$ dependence observed via spin-lattice relaxation measurements[22,27,43] (impractical herein given the long time scales under cryogenic temperatures). In particular, we find for $\rho \sim 4\times10^{13}$ Ohm a cutoff frequency $\omega_c \equiv \left(\sqrt{8/3}\,\epsilon\rho d\right)^{-1} \sim 170\times10^3$ rad·s⁻¹, consistent with the higher frequency range probed in these prior studies (0.1–1 MHz). Further, we calculate coherence lifetimes in the tens of microseconds, in qualitative agreement with our observations.

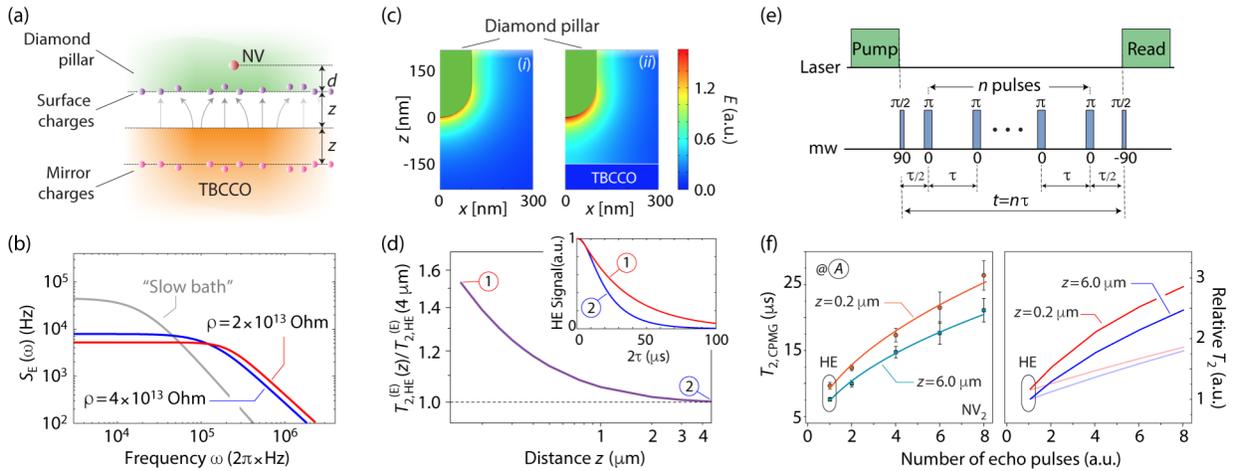

**Fig. 3 | Electric noise from surface charges.** (a) Charge carriers on the diamond surface lead to electric field fluctuations at the NV site; virtual image charges in the superconductor cancel the electric field component parallel to the TBCCO surface. (b) (Blue trace) Spectral density of the electric noise as a function of frequency for a surface resistivity $\rho = 4 \times 10^{13}$ Ohm. (Red trace) Proximity to the superconductor brings down the surface resistivity, thus leading to a change in $S_E(\omega)$. (Grey trace) Background Lorentzian noise insensitive to the superconductor; the amplitude and cutoff frequency are respectively $4 \times 10^4$ Hz and $2 \times 10^4$ rad·s$^{-1}$. (c) Calculated electric field amplitude for a diamond pillar (*i*) far and (*ii*) 150 nm away from the TBCCO film. Greater electric field amplitudes reveal the accumulation of surface charge when in proximity to the superconductor. (d) Relative change of the NV coherence lifetime in the presence of electric field fluctuations as a function of the tip distance to the superconductor. The insert shows the NV Hahn-echo envelope as calculated from Eqs. (1) and (2). (e) Schematics of the multi-pulse Carr-Purcell–Meiboom-Gill (CPMG-$n$) protocol. (f) Measured (left) and calculated (right) coherence lifetimes using the "two-source" noise model in (b). For illustration purposes, faint traces in the right plot show results with an electric-noise-only model with parameters chosen so as to match the observed Hahn-echo change. Solid lines in the left plot are guides to the eye.

Unlike the case for magnetic noise (SI, Sections V and VI), Eq. (2) shows no explicit *z*-dependent shielding, a counterintuitive result that stems from the superconductor-induced self-screening of surface carriers (SI, Section VII). The presence of image charges in the TBCCO film, however, effectively lowers the inter-carrier repulsion energy on the tip wall closest to the sample, which correspondingly boosts the local surface carrier density (Fig. 3c). In the low-frequency regime applicable here (i.e., where $\omega \lesssim \omega_c$ and $S_E(\omega)$ is frequency insensitive), the corresponding reduction of the diamond surface resistivity — inversely proportional to the carrier concentration — leads to a concomitant growth of the spin coherence lifetime, in qualitative agreement with our observations. This is explicitly shown in Fig. 3d where we combine Eqs. (1) and (2) to calculate the NV signal envelope under a Hahn-echo protocol; proximity to the superconductor halves the surface resistivity and leads to NV signal changes comparable to those seen in Fig. 2. Note that while a sheet of metal could arguably lead to similar effects, the magnetic[10,44] and electric[11,45] noise it produces — absent in a superconductor material[6-12] but strong in metallic systems — may counter any coherence lifetime gains.

Interestingly, electric noise does not seem to be single-handedly responsible for the observed NV spin dynamics. This is shown in Figs. 3e and 3f where we extend the observations of Fig. 2 to determine the NV transverse relaxation time in the case of multi-pulse spin control protocols. These measurements provide complementary information as pulse trains with greater number of inversion pulses tend to probe a higher frequency range (in turn, a consequence of the short inter-pulse spacing required for efficient spin decoupling). We find qualitative agreement with our observations only if we also include a second, "background" noise source insensitive to the superconductor (faint black trace in Fig. 3b). One such contribution could be the noise created by surface paramagnetic defects (e.g., in the form of dangling bonds); note that since carriers trapped on the diamond surface can also give rise to paramagnetic centers[35], our findings point to interwoven electric and magnetic (spin) noise sources of comparable impact.

**Toward $T_2$-weighted NV imaging**

The relation between NV coherence and superconductor proximity can, in principle, be exploited to reconstruct a $T_2$-based map of the film. Similar to known techniques in magnetic resonance imaging[46], this form of transverse spin relaxometry — based on the Hahn echo or tailored to multi-pulse trains — conveys valuable information, particularly because the timed structure of the protocol can be potentially adapted to expose dynamical processes not apparent through standard ODMR sensing[47]. In the present case, however, the implementation is challenging because supercurrents induced by the antenna on the TBCCO patch can screen or enhance the mw field as the NV probe moves from one point to the next. We gain a crude appreciation of the superconductor impact on the local mw through Fig. 4a, displaying the duration of the mw pulse required for NV spin inversion along a straight line partly overlapping with the superconductor film. After reaching a local minimum — stemming from mw-field enhancement near the edges — we measure long pulse durations as we enter the superconductor, indicating mw field shielding.

To circumvent this problem, we modify our protocol so as to scale the mw power at each point during the scan in a way that maintains unaltered the duration (and net spin rotation) of all pulses in the sequence; unlike the case where pulses change their length, this strategy guarantees NV spin control over a constant excitation bandwidth (see SI, Section VIII and Fig. S8). Further, we shift at each point the mw frequency — varying across the probed section, Fig. 4b — so as to ensure resonant NV spin manipulation at all positions (Fig. 4c). Figure 4d shows the result: Consistent with our prior findings, we observe a growth of the spin coherence lifetime when the NV hovers above the superconductor film (right section of the plot). Intriguingly, we also find a local reduction of $T_{2,HE}$ near the TBCCO edge, which roughly correlates with the observed mw and magnetic field minima across the same path. We hypothesize this change could stem from a surface charge redistribution away from the pillar as other sections of the cantilever — in the form of a 5 µm wide board, see Suppl. Fig. 2a — get close to the superconductor edge during the scan; additional work, however, will be

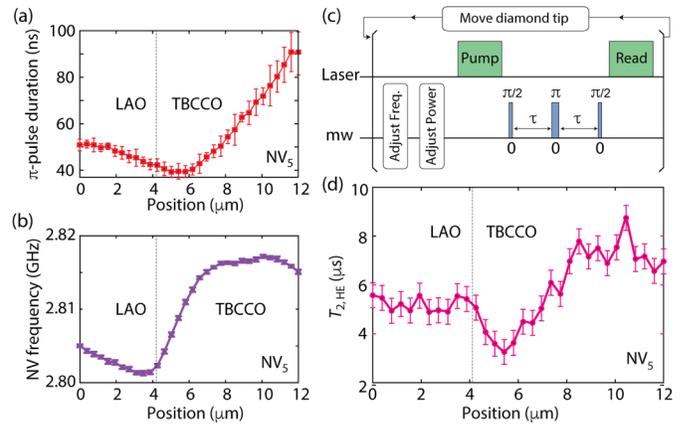

**Fig. 4 | One-dimensional imaging of NV transverse relaxation.** (a) Upon implementing a Rabi protocol (where we monitor the NV response after a mw pulse of variable length), we determine the π-pulse duration as the tip moves across the TBCCO boundary (vertical dotted line). (c) NV resonance frequency as a function of position along the one-dimensional path in (a). (c) $T_{2,HE}$ imaging protocol. (d) NV Hahn-echo coherence lifetime $T_{2,HE}$ along the linear path in (a) and (b); the NV distance to the superconductor is 150 nm. In (a) through (d), error bars indicate standard deviations.

## Discussion

In summary, our results show that proximity to a superconductor can extend the coherence lifetime of a shallow NV center. This behavior is reminiscent of prior work on cold atoms or ions trapped near metallic and superconducting electrodes[6-12], although the complex nature of the diamond surface makes the mechanisms at play herein especially hard to unravel. Our theoretical modeling suggests that electric noise plays a major role in the limit of high surface resistivity (the case for oxygen-terminated diamond[40,48]). Longer coherence lifetimes are possible thanks to a superconductor-induced boost in the local surface electron density. We warn, however, that a detailed microscopic modeling of a resistivity change in a strongly insulating system is difficult, implying that the proposed noise suppression process must be seen as tentative. Other possible mechanisms contributing to a change in surface resistivity include the drop of the carrier thermal activation gap from screening[49], and a screening-induced reduction of the disorder potential experienced by charge carriers.

Our results also reveal a close inter-dependence between electric and magnetic noise sources, one growing with the surface resistivity, the other with the surface conductivity (SI, Sections VI and VII). Further, analysis of the system response for different control protocols suggests spin noise — comparable in magnitude to the electric contribution but insensitive to the action of the superconductor — is also present. Spin and electric noise are likely related because an increase in the electron surface density should be accompanied by the enhanced depletion of donors in the NV vicinity (most notably, neutral nitrogen)[48,50]. While $N^0$ is paramagnetic, $N^+$ ions are spin-less, implying this class of process also reduces the magnetic noise (without the need for Meissner shielding).

The interplay between the superconducting sample, the spin probe, and the environment created by its crystal host serves as the basis for new imaging modalities. Besides the transverse relaxation measurements shown here, proximity to the superconductor also produces changes in the shape and overall structure of the echo revival pattern, particularly in cases where the NV strongly couples to an adjacent $^{13}C$ spin (not shown here for brevity). This form of contrast — tentatively attributed to Stark shifts produced by reconfigured donors around the NV[38] — could be advantageous as the required evolution time is comparatively shorter.

Looking forward, an interesting possibility is to exploit time-resolved NV magnetometry to derive the noise spectral density of the superconductor film from a comparison to a reference far away from its surface. This approach could help monitor thermally activated processes in type-II superconductors such as vortex creep and depinning[51], or spectroscopically characterize spin- and charge-induced noise in superconducting qubits[52]. In the latter case, nm-resolved decoherence imaging of Josephson junctions seems particularly attractive provided tip contamination stemming from close sample proximity — a problem herein, see SI, Section I — can be mitigated.

**Acknowledgments**. The authors acknowledge useful discussions with P. Ghaemi and T. Brick. R.M. acknowledges support from the National Science Foundation through grant NSF-1914945. Work by T.D and V.M.M. was supported by the National Science Foundation under grant NSF-QIITAQS-1936351. Work by N.P., Z.S., H.J., J.H., and P.R.Z. was supported by the National Science Foundation under grant NSF-1726573. Work by A.L., D.P., and C.A.M was supported by the U.S. Department of Energy, Office of Science, National Quantum Information Science Research Centers, Co-design Center for Quantum Advantage (C2QA) under contract number DE-SC0012704. All authors acknowledge access to the facilities and research infrastructure of the NSF CREST IDEALS, grant number NSF-HRD-1547830.

**Author contributions**. R.M. and T.D. conducted the experiments. R.M., N.P., Z.S., H.J., J.H., P.R.Z., A.L., D.P., T.A., V.M.M., and C.A.M. conceived and developed the microscope. I.S., T.D., and C.A.M. led the theoretical and numerical modelling with assistance from R. M. and P.D.E.; C.A.M. wrote the manuscript with input from all authors. V.M.M. and C.A.M. supervised the project.

**Competing interests**. The authors declare no competing interests.

**Correspondence and requests for materials** should be addressed to C.A.M.

# Supplementary material for

# "Spin dynamics of a solid-state qubit in proximity to a superconductor"

Richard Monge[1,2], Tom Delord[1], Nicholas Proscia[1], Zav Shotan[1], Harishankar Jayakumar[1], Jacob Henshaw[1], Pablo R. Zangara[1], Artur Lozovoi[1], Daniela Pagliero[1], Pablo D. Esquinazi[3], Toshu An[4], Inti Sodemann[5,6], Vinod M. Menon[1,2], Carlos A. Meriles[1,2,*]

**I. Experimental setup**

Figure S1 displays a block diagram of our instrument, a scanning probe-assisted confocal/ODMR microscope adapted to a closed-cycle cryo-workstation (Montana Instruments). Two independent nano-positioning stacks within the workstation chamber are used to vary the relative location of the sample and atomic force microscopy (AFM) probe. Optical excitation is carried out with a continuous-wave (cw) laser at 532 nm and an acousto-optic modulator (AOM) to generate light pulses with ~10 ns temporal resolution. We use a 0.7-NA, 3.2-mm- working-distance objective for sample and/or scanning probe illumination and photon collection; although within the cryo-workstation main chamber, the objective sits on the thermally insulated, not-cryo-cooled section of the platform, thus remaining near room temperature at all times. Under normal operation, fluorescence imaging is attained by scanning the sample while both the laser beam and the cantilever remain fixed.

Except for the objective, all control stages sit on an actively-cooled, floating breadboard featuring vibration isolation technology with a very low resonance frequency that effectively transforms rapid vibrations into very slow displacements. As a result, the distance between the tip and the sample surface remains virtually unchanged at all times, with relative displacements below 1 nm. Integrated control of all capabilities — including sample/tip positioning, magnetic resonance protocol design, and optical/atomic force microscopy — is attained via a home-made user interface.

Throughout the present experiments, we use commercial all-diamond scanning probes (Qnami), each hosting a few NV centers derived from $^{14}$N implantation; these probes have the geometry of a "diving board" with a protruding, 200-nm-wide cylinder hosting a few NVs ~10 nm from the scanning surface[53] (Fig. S2a). The spin probe fluorescence is coupled into a single mode fiber (also serving as the confocal pinhole, not shown) and detected with the help of a dichroic mirror and two avalanche photo-detectors (APD) in the Hanbury-Brown-Twiss geometry; a time-correlated photon counting module is used to establish photon correlations when required.

We implement NV ODMR through the use of a microwave signal generator (Rohde-Schwartz SMB100A)

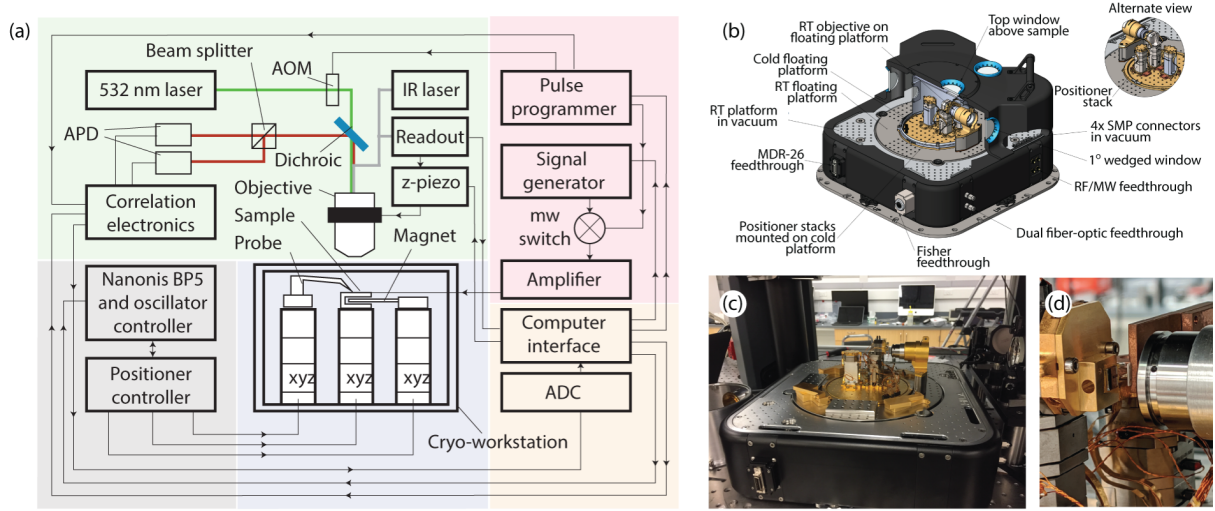

**Fig. S1 | Confocal/ODMR/atomic force microscope**. (a) Block diagram of the AFM/confocal microscope. Optics in green; magnetic resonance in red; scanning probe in grey; cryo in blue; control in orange. Not shown for simplicity is the feedback mechanism to adjust the laser focal point upon displacements of the workstation base plate. APD: Avalanche photo-detector. AOM: Acousto-optic modulator. ADC: Analog-to-digital. XYZ: Nanopositioners/scanners. (b) Three-dimensional rendering of the cryo-workstation. (c,d) Blown-up views of the cold chamber and sample holder section of the microscope.

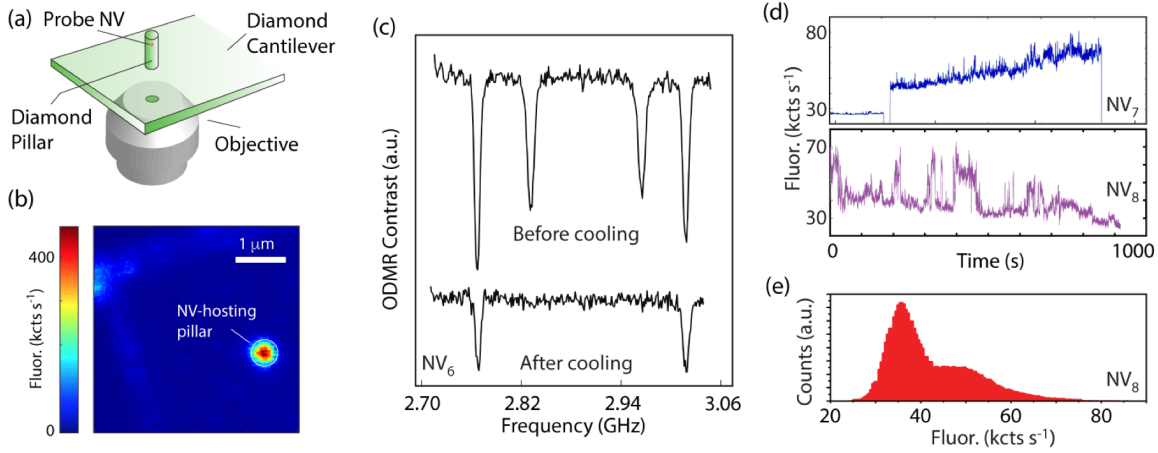

**Fig. S2 | Characterization of the probe NVs.** (a) We use confocal microscopy to monitor a commercial scanning probe with the NV sitting near the surface of a pillar protruding off the diamond cantilever. (b) Confocal fluorescence of a typical diamond probe at room temperature; the faint fluorescence at the edges demarks the physical edges of the cantilever. (c) ODMR spectra from two differently-oriented NVs at the tip before and after cooling from room temperature to 80 K (vertically displaced for clarity); the ODMR resonance dip from one of the NVs disappears and the signal contrast in the other diminishes to about a third. We also observe activation of background emitters throughout the cantilever, not shown for brevity. (d) Vacuum conditions often render the probe NV unstable, sometimes leading to gradual fluorescence growth and sudden bleaching or blinking (upper and lower panels, respectively). (e) Histogram of the fluorescence emission in the lower panel of (d) indicating a bimodal distribution, likely the result of local charge instabilities activated by the laser beam.

and a 30 W broadband amplifier (Mini-Circuits ZHL-16W-43-S+). For time-resolved measurements, we use a 3-ns-risetime switch to generate microwave pulses, which we deliver here with the help of a 25-μm-diameter copper wire overlaid on the sample surface. Unless otherwise noted, we monitor the $|m_S = 0\rangle \leftrightarrow |m_S = -1\rangle$ transition. The optical contrast in these NVs typically reaches optimal values (12-15% for ODMR measurements) with photon counts of up to 400 Kcts·s$^{-1}$ under saturation conditions thanks to preferential wave-guiding by the diamond pillar (Fig. S2b). The magnetic field from a permanent magnet allows us to spectroscopically separate NVs with different orientations. Figure S2c shows an example where individual NVs along two crystallographic axes can be seen.

We observe large variations in the Hahn-echo lifetime $T_{2,HE}$, ranging from ~5 μs to 75 μs depending on the tip we use. Further, we find NVs in these tips exhibit lower fluorescence at low temperatures and are prone to blinking and/or sudden quenching (especially below ~50 K, see Figs. S2d and S2e), a major complication that translates into frequent tip replacement. We hypothesize this problem stems from charge instabilities exacerbated under vacuum and cryogenic conditions.

## II. Sample characterization

The superconductor sample is a 500-nm-thick film of TBCCO 2212 on a 0.4 mm lanthanum aluinate (LAO) substrate[54]. The sample was manufactured by DuPont and kept in dry storage since fabrication. As shown in Fig. S3a, the superconductor substrate was diced into 5×5 mm² squares and patterned through optical lithography with negative resist (NR9-1000P Futurex Inc) and wet etching[55] (0.5 mol L$^{-1}$ citric acid aqueous solution for 45 min, followed by a 5 s wash in an alkaline solution RD6 Futurex Inc which was found to lower re-deposition). The sample was kept under atmospheric pressure for weeks before use.

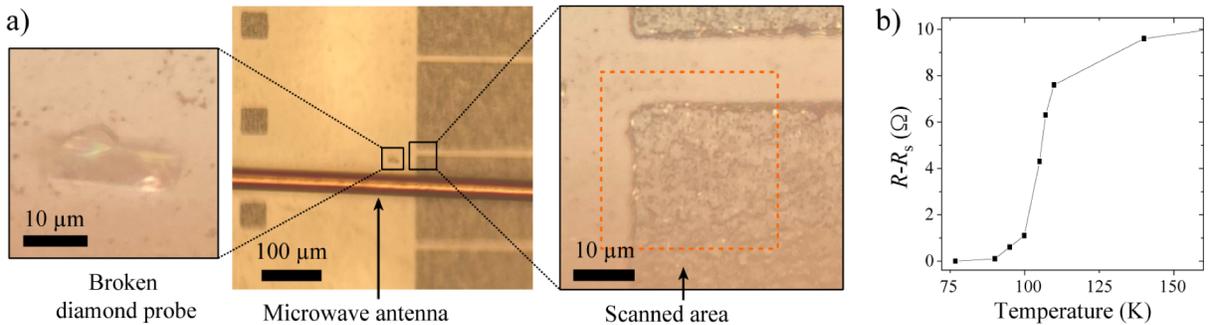

**Fig. S3 | TBCCO sample characteristics.** (a) White light microscope images of the sample also showing the antenna and a broken diamond probe. The dark (white) patches are the superconductor (TBCCO) and the bare substrate (LAO). (b) Resistance of a 5×5 mm² measured with two leads around the transition temperature.

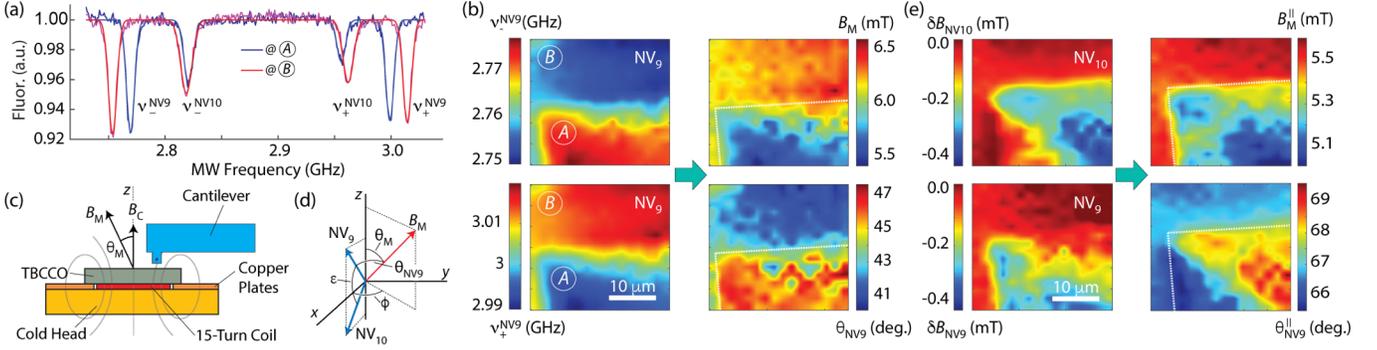

**Fig. S4 | Vector magnetometry.** (a) ODMR of a diamond tip; super-imposed solid lines are Gaussian fits. We identify two NVs aligned along inequivalent crystal axes, which we use to determine $\mathbf{B}_M$. (b) ODMR imaging of a corner section of the TBCCO film as determined from monitoring $\nu_-^{NV9}$ and $\nu_+^{NV9}$ (upper and lower left-hand side images, respectively). From these measurements and Eqs. (S1) and (S2), we derive the field amplitude $B_M$ and angle $\theta_{NV9}$ relative to $NV_9$ (upper and lower right-hand side images, respectively). Points (A) and (B) on the left-hand side maps indicate the positions corresponding to the ODMR traces in (a). (c) In addition to the external field $\mathbf{B}_M$ from a permanent magnet, we use a small coil aligned with the sample substrate to produce a magnetic field $\mathbf{B}_C$ of known amplitude and perpendicular to the substrate. (d) In the absence of a superconductor, $\mathbf{B}_M$ has an amplitude of $B_M^{(0)} = 6.54 \pm 0.05$ mT and forms an angle $\theta_M^{(0)} = 43.7 \pm 3.6$ deg. with the substrate normal. (e) ODMR imaging of $\delta B_{NVj} = B_{NVj} - B_{NVj}^{(0)}$ for $NV_j$, $j = 9, 10$ across the same TBCCO section (upper and lower left-hand side images, respectively), which we then use to derive the amplitude $B_M^{\parallel}$, corresponding to the projection of $\mathbf{B}_M$ onto the plane formed by $NV_9$ and $NV_{10}$ (upper right-hand side image). We also determine the angle $\theta_{NV9}^{\parallel}$ formed by this projection and the $NV_9$ axis (lower right-hand side image). All experiments are carried out at 69 K and the tip distance to the surface is 150 nm; the dashed white contour in all magnetometry images indicates the boundaries of the TBCCO patch.

Resistance measurements provided by the seller gave a surface resistance below $3 \times 10^{-4}$ Ω at 80 K. In Fig. S3b, an estimation of $\mathcal{T}_c$ was performed by measuring the resistance of a 5×5 mm² square sample with two leads wire-bonded to the sample. A sharp drop of the resistance is visible between 90 K and 110 K corresponding to the superconducting transition. We attribute the residual resistance $R_s = 23$ Ω to the poor contact between the superconducting film and the measurement leads.

We have observed that close tip proximity to the TBCCO surface leads to fast degrading of the NV fluorescence, which first becomes unstable and ultimately decays to negligible values. We hypothesize this behavior derives from tip contamination, though we presently ignore the mechanisms at play. To extend the tip lifetime, all experiments herein are limited to working distances of 150 nm or more.

Magnetic fields whose component perpendicular to the surface exceeds values of order ~10 mT are seen to quench the superconducting state, the reason why we limit our measurements to lower field amplitudes. Throughout our experiments, we typically cool down the sample below $\mathcal{T}_c$ after applying the external magnetic field. Zero field cooling, however, produced no observable changes in the TBCCO response; the latter includes the magnetometry measurements discussed immediately below.

### III. Vector reconstruction of the external magnetic field

Since the relative orientation of the applied magnetic field can influence the NV spin coherence lifetime[56], we make use of ODMR magnetometry to map the impact of the TBCCO film on $\mathbf{B}_M$. In the experiments of Fig. S4, we choose an NV center — here denoted $NV_9$ — whose crystallographic orientation coincides with that of $NV_1$ in Fig. 1 of the main text. The magnetic field magnitude $B_M$ and polar angle $\theta_{NV9}$ relative to the axis of $NV_9$, can be obtained from the $m_S = 0 \leftrightarrow m_S = \pm 1$ transition frequencies — respectively denoted $\nu_-$ and $\nu_+$, see Fig. S4a — using the relations[57]

$$B_M^2 = \frac{h^2}{3\mu_B^2 g^2}(\nu_-^2 + \nu_+^2 - \nu_-\nu_+ - D^2)\bigg|_{@NV9} \quad (S1)$$

and

$$\theta_{NV9} = \frac{1}{2}\text{acos}\left(\frac{f(\nu_-,\nu_+) - g(\nu_-,\nu_+)}{h(\nu_-,\nu_+)}\right)\bigg|_{@NV9} \quad (S2)$$

where

$$f(\nu_-,\nu_+) \equiv 7D^3 + 2(\nu_- + \nu_+)(2(\nu_-^2 + \nu_+^2) - 5\nu_-\nu_+),$$
$$g(\nu_-,\nu_+) \equiv 3D(\nu_-^2 - \nu_-\nu_+ + \nu_+^2),$$
$$h(\nu_-,\nu_+) \equiv 9D(\nu_-^2 - \nu_-\nu_+ + \nu_+^2 - D^2),$$

and $D = 2.877$ GHz is the NV zero field splitting at 69 K. Figure S4b shows the measured frequencies for $NV_9$ across the TBCCO area of interest (left-hand side plots) and the corresponding maps of the field amplitude and angle relative to $NV_9$ (respectively, upper and lower images on the right-hand side). We find a ~20% reduction of the magnetic field amplitude above the TBCCO film, a finding consistent with prior observations in similar samples[58]. Further, we observe only a minor change of the field direction (not exceeding ~8 deg. in the region of the TBCCO patch), which allows us to conclude that the increase of the NV coherence lifetimes cannot stem from superconductor-induced changes in $\mathbf{B}_M$.

A closer inspection of the magnetometry images in Fig. S4b shows they do not accurately reproduce the rectangular geometry of the TBCCO patch (dashed white line), a consequence of the limited sensitivity of Eqs. (S1) and (S2) to small field changes. Fortunately, however, the diamond tip used for these experiments hosts NVs with two different crystallographic orientations (Fig. S4a), implying that additional information can be attained by simultaneously monitoring the response of both NVs. One possibility is to make use of Eqs. (S1) and (S2) to find the angle $\theta_{NV10}$ between $\mathbf{B}_M$ and $NV_{10}$. We also resort to the magnetic field $\mathbf{B}_C$ from an ancillary coil below the TBCCO film (Fig. S4c) to determine the orientation of the plane containing $NV_9$ and $NV_{10}$ relative to the sample surface. From these constraints, we find that in the absence of the superconductor, $\mathbf{B}_M$ forms an angle $\theta_M^{(0)} \approx 43$ deg. with the sample surface; by the same token, the angle relative to the plane formed by $NV_9$ and $NV_{10}$ — perpendicular to the sample surface, see Fig. S4d — is $\phi_M^{(0)} \approx 32$ deg.

Alternatively, we can combine the frequency shifts experienced by $NV_9$ and $NV_{10}$ to reconstruct the projection $\mathbf{B}_M^{\parallel}$ on the plane they form. Denoting as $B_{NV9}$ ($B_{NV10}$) the projected magnitude of $\mathbf{B}_M$ along the $NV_9$ ($NV_{10}$) axis, one can show that

$$B_M^{\parallel} = \sqrt{B_{NV9}^2 + \left(\frac{B_{NV10} + B_{NV9}\cos\varepsilon}{\sin\varepsilon}\right)^2}, \quad (S3)$$

and

$$\theta_{NV9}^{\parallel} = \left(\frac{\pi}{2} - \operatorname{atan}\left(\frac{B_{NV9}\sin\varepsilon}{B_{NV10} + B_{NV9}\cos\varepsilon}\right)\right), \quad (S4)$$

where $\varepsilon$ is the angle between the two NV axes, here considered to be 70.5 deg., and $\theta_{NV9}^{\parallel}$ denotes the angle between $\mathbf{B}_M^{\parallel}$ and $NV_9$. Fig. S4e shows the results: The projected field amplitude (upper right-hand side image) recaptures the rectangular geometry of the TBCCO patch, including the attenuated Meissner shielding of $\mathbf{B}_M$ near the boundaries[59]. We warn, however, that the above equations consider $\nu_+$ and $\nu_-$ symmetrically distributed about the midpoint — a crude approximation in the present case — and must therefore be seen as approximate.

Lastly, we observe no substantial changes in the measured magnetometry images if we increase the tip separation to a few microns (not shown here for brevity). This observation indicates that the field varies comparatively slowly as a function of distance, and hence does not correlate with the observed changes in the NV spin coherent response.

### IV. Ramsey measurements

Whereas Hahn-echo and CPMG measurements yield

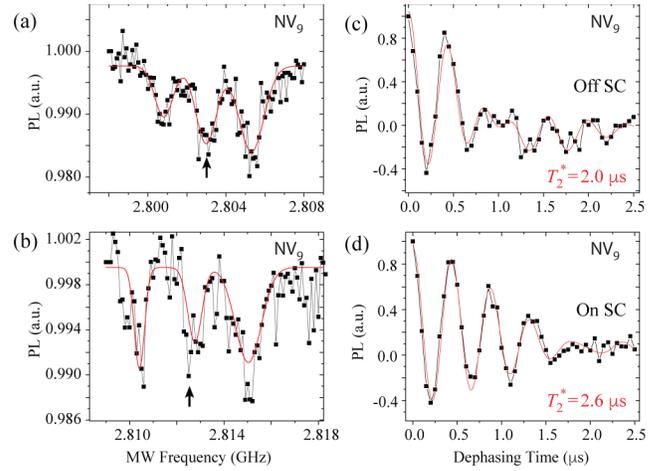

**Fig. S5 | Effect of the superconductor on the Ramsey response.** (a) Low power ODMR spectra (a) away from and (b) above the superconductor. Ramsey oscillations (c) away from, and (d) above the superconductor. Red traces indicate fits. The black arrows in (a) and (b) indicate the mw frequencies used.

information on the magnetic noise spectrum over select frequency bandwidths, we can also probe the overall magnetic noise amplitude via the dephasing time under free evolution[60], $T_2^*$. We perform $\pi/2 - \tau - \pi/2$ Ramsey sequences at two locations, namely, away from and on the superconductor at a vertical distance of 150 nm. We follow the same procedure as in the echo sequences: At each location, we first evaluate the spin resonance and Rabi frequency; to mitigate slow instrumental fluctuations, Ramsey measurements are performed at a frequency close to (but slightly away from) resonance. Due to the NV hyperfine coupling to the native $^{14}$N nuclear spin, Ramsey oscillations occur at three different frequencies corresponding to nuclear spin projections $m_I = 0, \pm 1$. To model the NV response, we first perform for each measurement a lower power ODMR as shown in Figs. S5a and S5b. The relative position of each peak is subsequently used to set parameter bounds on the Ramsey signal fit. Figures S5c and S5d show the Ramsey oscillations measured off and on the superconductor along the corresponding fits. We obtain $T_2^*$ values of $2.6 \pm 0.4$ μs and $2.0 \pm 0.16$ μs respectively on and off the superconductor, a 1.5-fold enhancement consistent with the Hahn-echo and CPMG observations in the main text.

### V. Meissner shielding of spin noise

The coherence of near-surface NV spins has been shown to be impacted by fluctuating magnetic fields created by surface paramagnetic impurities[21,22]. Close to the superconductor, the Meissner effect tends to suppress these fluctuations, therefore enhancing the NV spin coherence. As presented in Fig. S6a, we consider an ideal superconductor generating a Meissner field equal to the field generated by the mirror image of the source spin impurity relative to the superconductor interface assuming no penetration depth[61]. For an uncorrelated spin bath, the magnetic field noise spectral density can be written as[21,62]

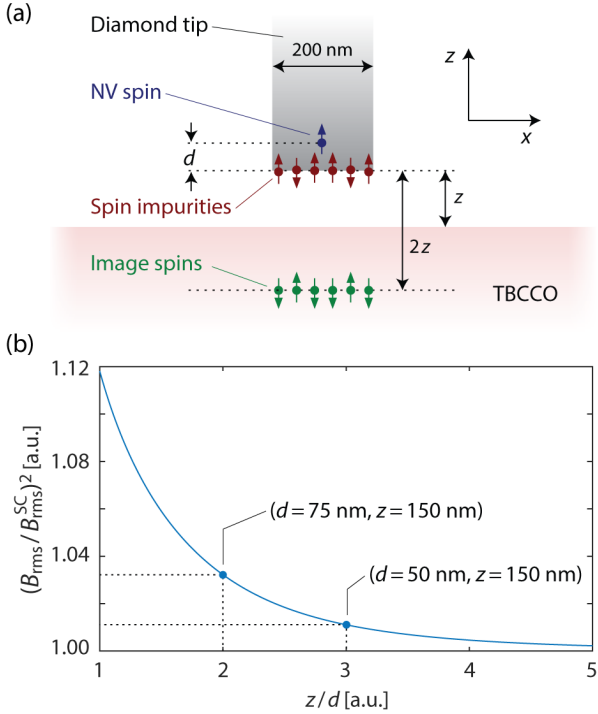

**Fig. S6 | Meissner shielding of spin noise.** (a) Diamond nano-pillar hosting an individual NV spin probe and an ensemble of surface paramagnetic impurities uniformly distributed over the pillar surface. The Meissner response of the superconductor can be approximated by a collection of image spins. (b) Suppression factor of the root mean square magnetic field generated by the spin bath as a function of $z/d$. For a given tip distance to the surface ($z = 150$ nm in the example), improved suppression is attained for deeper NVs.

$$S(\omega) = \sum_k B_k^2 \frac{\gamma_k}{\gamma_k^2 + \omega^2/4}, \quad (S5)$$

where $1/\gamma_k$ and $B_k$ respectively denote the correlation time and field created by the $k$-th spin. For a bath with homogeneous correlation time, the noise at a given frequency is proportional to the mean square magnetic field:

$$B_{\text{rms}}^2 = \sum_k B_k^2. \quad (S6)$$

The effect of the Meissner effect on coherence is therefore obtained by calculating the ratio of the mean square field with and without the superconductor. For simplicity, we consider the NV spin and the bath spin impurities aligned along the $z$-axis, with the spin impurities homogeneously distributed on the surface of a cylindrical diamond pillar; we also assume the continuous limit, neglecting the effects of discrete spin impurity sites.

Denoting $B_{\text{rms}}^{\text{SC}}(z/d)$ the root-mean-square spin noise field in the presence of the superconductor, we show in Fig. S6b the suppression factor, $B_{\text{rms}}^{\text{SC}}/B_{\text{rms}}$, as a function of the relative distance to the sample surface $z/d$. Stronger suppression of the noise is observed when the NV lies further away from the spin impurities, which favors a better screening by the superconductor. The tip-surface distance, $z$, in our experiments typically ranges from 150 to 350 nm. On the other hand, the NV spin probes we use have implantation energies of 6 and 12 keV. For those energies, simulations put the NV depth in the 5-13 nm and 11-23 nm ranges, respectively, but experiments performing AFM scans with the same commercial tips have found NV distances in the 60-79 nm range[53]. For a perfect superconductor, we therefore obtain a coherence enhancement due to Meissner shielding ranging from 0.003% (9 nm deep NV) to 3.5% at best (79 nm deep NV), more than one order of magnitude weaker than the observed effect.

## VI. Meissner shielding of current-induced magnetic noise

In this section, we discuss the contribution to the local magnetic noise at the NV site created by orbital electric currents on the diamond surface (to which we assign a surface conductivity $\sigma$). The model presented in Fig. S7a is based on a picture analogous to that shown in the main text: Current fluctuations give rise to magnetic field noise at the NV center via the Biot-Savart law. We model the superconductor as an ideal "current shield" meaning that for any physical current on the diamond surface there is an equal and opposite virtual current inside the superconductor at a distance $2z$ from the tip surface, hence guaranteeing a vanishing normal magnetic field at the TBCCO surface. While the above is a crude approximation that ignores the finite penetration depth in the superconductor, our calculation can be seen as the best-case scenario for the purposes of noise reduction.

We follow closely the derivations of Refs. [63] and [64] (see also Ref. [65] for a related discussion). We focus for concreteness on the fluctuations of the component of the magnetic field along the $z$-direction. Then Eq. (G12) of Ref. [64] can be rewritten as

$$B_z(\mathbf{q}, t) = -i\frac{\mu_0}{2}\left(e^{-|q|d} - e^{-|q|(d+2z)}\right) j_\perp(\mathbf{q}, t), \quad (S7)$$

where the first term is the contribution to the field from a transverse current $j_\perp$ fluctuation with wave-vector $\mathbf{q}$ on the diamond surface, and the second one is the contribution from the virtual image current in the superconductor. Following a similar analysis, one arrives at a slightly modified version of Eq. (G25) in Ref. [64], relating the magnetic field response function $\chi_{B_z,B_z}(\omega)$ to the transverse conductivity of the diamond surface

$$\chi_{B_z,B_z} = \frac{\mu_0^2 \hbar \omega}{4} \int \frac{d^2\mathbf{q}}{(2\pi)^2} \left(e^{-|q|d} - e^{-|q|(2z+d)}\right)^2 \sigma_\perp. \quad (S8)$$

In general, the transverse conductivity $\sigma_\perp$ can have a non-trivial dependence on wave-vector and frequency, but since in our case the NV center is located at a distance larger than the mean-free path and we probe frequencies much smaller than the scattering rate, it is legitimate to express $\sigma_\perp$ as the frequency and wave-vector-independent Drude conductivity, i.e., $\sigma_\perp \approx \sigma$. Upon performing the integral in Eq. (S8), we write the magnetic field noise as

$$S_B = \frac{\gamma_e^2 \mu_0^2 k_B \mathcal{T}}{32\pi} \frac{\sigma(z)}{d^2}\left(1 - \frac{2d^2}{(z+d)^2} + \frac{d^2}{(2z+d)^2}\right), \quad (S9)$$

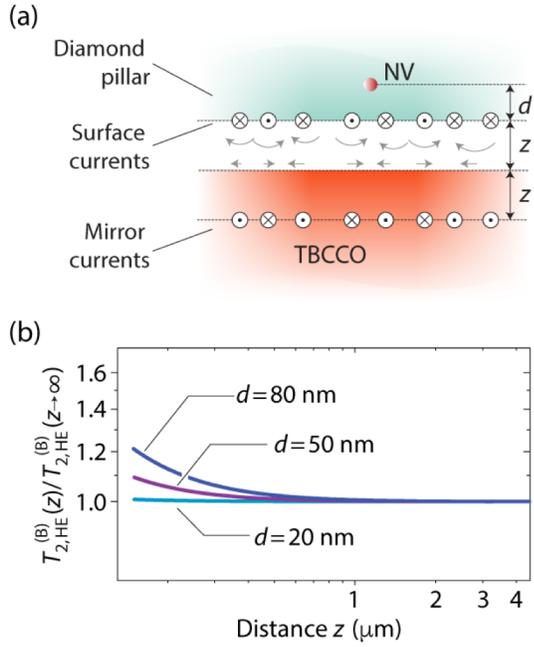

**Fig. S7 | Shielding of magnetic noise from surface currents.** (a) The diamond tip hosts a collection of moving charges near the surface leading to magnetic field fluctuations. Proximity to the TBCCO film can be crudely modeled through image currents cancelling the component of the magnetic field noise perpendicular to the superconductor surface. (b) Calculated relative change of the coherence lifetime in the presence of magnetic fluctuations as a function of the tip distance to the TBCCO surface for three different NV depths.

where we used the NV spin gyromagnetic coupling $\gamma_e$ and the fluctuation dissipation theorem to express the spectral density in terms of the magnetic response function, i.e., $S_B = \gamma_e^2 \chi_{B_z,B_z} \coth(\hbar\omega/2k_B\mathcal{T})/2$; Eq. (S9) results after taking the high-temperature limit.

Figure S7b shows the calculated ratio between the Hahn-echo coherence lifetimes in the presence of magnetic noise, $T_{2,\mathrm{HE}}^{(B)}(z,d)/T_{2,\mathrm{HE}}^{(B)}(z\to\infty,d)$ for different NV depths (including the change that derives from charge accumulation at the tip, see main text). We find that the fractional change of the transverse spin lifetime trends to values lower than those observed. Further, the low surface conductivity of oxygen-terminated diamond[40-42] — of order $10^{-10} - 10^{-17}$ Ohm$^{-1}$ — yields relaxation rates $\left(T_{2,\mathrm{HE}}^{(\mathrm{M})}\right)^{-1}$ of order $10^{-6} - 10^{-14}$ s$^{-1}$, several orders of magnitude below the experimental results. While uncertainty in the composition of the diamond surface makes absolute comparisons difficult, the large discrepancy allows us to rule out superconductor-induced shielding of magnetic interactions as the leading mechanism in our observations.

### VII. Electric noise in proximity to a superconductor

Here, we discuss the contribution to the local electric noise at the NV center created by charge density fluctuations on the diamond surface. For our purposes, the TBCCO film can be crudely seen as an ideal metal whose action can be modeled through image charges, instantaneously cancelling the electric field component parallel to the superconductor surface (Fig. 4a). By solving the electrostatic Poisson equation, one arrives at the counterpart to Eq. (S7) relating the normal electric field to charge density fluctuations $\rho(\mathbf{q},t)$ of wave-vector $\mathbf{q}$ on the diamond surface

$$E_z(\mathbf{q},t) = \frac{1}{2\varepsilon}\left(e^{-|q|d} - e^{-|q|(d+2z)}\right)\rho(\mathbf{q},t). \quad (S10)$$

For simplicity, Eq. (S10) assumes the system is embedded in a medium of uniform dielectric constant $\varepsilon$. Following an analysis similar to that in Ref. [64], one arrives at the counterpart to Eq. (S8), namely

$$\chi_{E_z,E_z} = \int \frac{d^2\mathbf{q}}{(2\pi)^2} \frac{\left(e^{-|q|d} - e^{-|q|(2z+d)}\right)^2}{4\varepsilon^2} \chi'_{\rho\rho}(\mathbf{q},\omega). \quad (S11)$$

Here, $\chi'_{\rho\rho}(\mathbf{q},\omega)$ is the imaginary (dissipative) part of the charge density susceptibility of the diamond surface. We model this response function assuming a two-dimensional hydrodynamic system based on Ohmic conduction. The charge density obeys the following equations

$$\partial_t \rho + \nabla \cdot j = 0, \quad (S12)$$

$$\partial_t j + \Gamma j = D\left(E^{\mathrm{self}} + E^{\mathrm{ext}}\right), \quad (S13)$$

where $D = \eta e^2/m_0$ is the Drude weight for carriers of concentration $\eta$, charge $e$, and mass $m_0$, $\Gamma$ is the carrier scattering rate, $E^{\mathrm{self}}$ is the electric field generated by the fluctuating charge density itself, and $E^{\mathrm{ext}}$ is an external perturbing electric field introduced for purposes of computing the response functions. After Fourier transforming, these electric fields can be written as

$$E^{\mathrm{self}} = iq V_q^{\mathrm{Coul}} \delta\rho_q, \quad (S14)$$

$$E^{\mathrm{ext}} = -i\,\mathbf{q}\phi_q^{\mathrm{ext}}, \quad (S15)$$

where $V_q^{\mathrm{Coul}}$ is the two-dimensional Fourier transform of the screened Coulomb potential given by

$$V_q^{\mathrm{Coul}} = \frac{(1 - \exp(-2qz))}{2\varepsilon q}. \quad (S16)$$

By solving the above equations, one obtains

$$\delta\rho_q = \frac{1}{\hbar}\chi_{\rho,\rho}(q,\omega)\phi_q^{\mathrm{ext}}, \quad (S17)$$

$$\chi_{\rho,\rho}(q,\omega) = -\frac{\hbar}{\frac{i\omega\rho}{q^2} + V_q^{\mathrm{Coul}}}. \quad (S18)$$

Here, $\rho = m_0(i\omega + \Gamma)/(\eta e^2)$ is the Drude resistivity of the diamond surface. At frequencies below the plasma resonance, the resistivity can be approximated as real, and we can rewrite the imaginary part as

$$\chi'_{\rho,\rho}(q,\omega) \approx \frac{\hbar\omega\rho}{\left(qV_q^{\mathrm{Coul}}\right)^2\left(1 + \left(\omega\rho/q^2 V_q^{\mathrm{Coul}}\right)^2\right)}. \quad (S19)$$

Replacing the above in Eq. (S11), we find

$$\chi_{E_z,E_z} = \hbar\omega\rho \int \frac{d^2\mathbf{q}}{(2\pi)^2} \frac{e^{-2|q|d}}{1 + \left(\omega\rho/q^2 V_q^{\mathrm{Coul}}\right)^2}. \quad (S20)$$

In the limit where $z \gg d$, we can neglect the screening correction in the Coulomb term in the denominators above, and we can also write $q \sim 1/d$ in the second term, i.e.,

$$\chi_{E_z,E_z} \sim \hbar\omega\rho \frac{1}{1+\left(\sqrt{8/3}\,\epsilon\omega\rho d\right)^2} \int \frac{d^2\mathbf{q}}{(2\pi)^2} e^{-2|q|d}, \quad (S21)$$

$$\chi_{E_z,E_z} \sim \frac{\hbar\omega\rho}{8\pi d^2} \frac{1}{\left(1+\left(\sqrt{8/3}\,\epsilon\omega\rho d\right)^2\right)}. \quad (S22)$$

The expression in Eq. (2) in the main text follows from that above after including the dipole moment coupling of the NV and accounting for the $\coth(\hbar\omega/2k_B\mathcal{T})/2$ factor. The NV spin coherence time can be then extracted from Eq. (1) using the filter function formula[36]

$$F_n(\omega t) = \left| 1 + (-1)^{n+1} e^{i\omega t} + 2 \sum_{k=1}^{n} (-1)^k e^{i\omega t_k} \cos(\omega \tau_\pi/2) \right|^2, \quad (S23)$$

where $t_k$ is the time of the $k$-th pulse in the train, and $\tau_\pi$ is the pulse duration.

### VIII. Microwave recalibration protocol for $T_2$ imaging

Heterogeneities in the amplitude and direction of the applied mw and magnetic fields — present in our system at temperatures below $T_c$ — complicate the implementation of sample scans based on the sequential application of pulsed NV control sequences. To circumvent this problem, we implement a Rabi protocol at each location, which we then use to determine the local duration of the mw π-pulse (Fig. S8a). We also make use of our ODMR magnetometry images — literally, mapping the NV $m_S = 0 \leftrightarrow m_S = -1$ transition frequencies at all positions throughout the scanned area — to maintain the mw excitation on-resonance irrespective of the tip position. During the scan, we build on this information to adjust the mw frequency and amplitude so as to ensure that the length of each pulse in the protocol remains unchanged (Fig. S8b). As the pulse length inversely relates to the excitation bandwidth, this strategy eliminates the signal distortions otherwise arising from the use of longer pulses with nominally equivalent spin rotations.

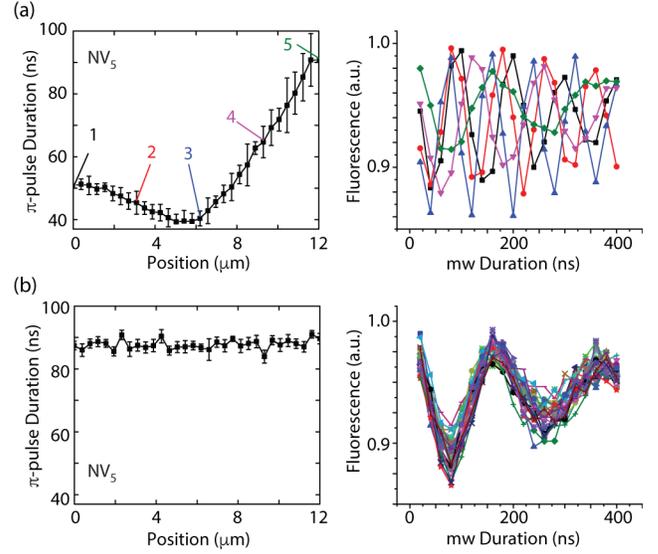

**Fig. S8 | Correction of mw amplitude during a $T_2$ scan.** (a) Duration of the mw π-pulse along a line-scan across the TBCCO patch boundary (see Fig. 4). Colored numbers indicate the locations corresponding to the Rabi plots on the right. (b) Same as in (a) upon automated recalibration of the mw power. The plot on the right reproduces the measured NV Rabi response at each location across the scan (32 sites in total).

### Data availability

The data supporting the findings of this study are available from the corresponding author upon reasonable request.